\documentclass[12pt]{article}

\usepackage{amsmath,amssymb,amsthm,graphicx,latexsym}
\usepackage{hyperref}
\usepackage{graphics}
\usepackage{setspace}
\usepackage{xcolor}
\newcommand{\ed}{\end{document}}
\newcommand{\beq}{\begin{equation}}
\newcommand{\eeq}{\end{equation}}
\newcommand{\beqa}{\begin{eqnarray}}
\newcommand{\eeqa}{\end{eqnarray}}
\newcommand{\bc}{\begin{center}}
\newcommand{\ec}{\end{center}}

\newcommand{\ba}{\begin{array}}
\newcommand{\ea}{\end{array}}
\newcommand{\pa}{\partial}

\newcommand{\de}{\delta}
\newcommand{\brr}{({\bf r}-{\bf r}')}
\begin{document}

\title{{\bf{Divergence Anomaly and Schwinger Terms:\\
Towards a Consistent Theory of  Anomalous  Classical Fluid
 }}}
	\author{  {\bf {\normalsize Arpan Krishna Mitra$^1$}$
			$\thanks{E-mail: arpankrishnamitra@hri.res.in}}$~$
		{\bf {\normalsize Subir Ghosh$^2$}$
			$\thanks{E-mail: subir\_ ghosh2@rediffmail.com}},$~$
		\\{\normalsize $^1$Harish Chandra Research Institute, HBNI,}
		\\{\normalsize Chhatnag Rd, Jhusi, Uttar Pradesh 211019}
		\\\\
		{\normalsize $^2$Indian Statistical Institute}
		\\{\normalsize  203, Barrackpore Trunk Road, Kolkata 700108, India}
		\\[0.3cm]
	}
	\date{}
	\maketitle
	
\begin{abstract} 
Anomaly, a generic feature of relativistic quantum field theory, is shown to be present in non-relativistic classical ideal fluid. A new result is the presence of anomalous terms in current algebra, an obvious analogue of Schwinger terms present in quantum field theory. We work in Hamiltonian framework where Eulerian dynamical variables obey an anomalous algebra (with Schwinger terms) that is inherited from modified Poisson brackets, with Berry curvature corrections, among Lagrangian discrete coordinates. The divergence anomaly appears in the Hamiltonian equations of motion. A generalized form of fluid velocity field can be identified with the "anomalous velocity" of Bloch band electrons appearing in quantum Hall effect in condensed matter physics. We finally show that the divergence anomaly and Schwinger terms satisfy well known Adler consistency condition. Lastly we mention possible scenarios where this new anomalous fluid theory can impact.
\end{abstract}

{\bf{Introduction:}} 
Anomalies are universal and thoroughly researched phenomena, present in  generic relativistic quantum field theories (QFT) (for reviews see \cite{tr}). It is a form of breakdown of certain conservation laws and corresponding symmetries in QFT that are present in the corresponding classical field theory. Naive application of  equations of motion on certain composite dynamical variables ({\it{e.g.}}  fermion current $\bar\psi(x)\gamma_\mu\psi(x)$) in a classical framework generates the conservation laws following Noether's principle. To give the iconic example of axial or Adler Bell Jackiw anomaly \cite{abj}, there are  vector and axial vector current conservation laws in   mass-less QED,  $\partial^\mu J^V_\mu(x)=\partial^\mu(\bar\psi(x)\gamma_\mu\psi(x))=0, \partial^\mu J^A_\mu(x)=\partial^\mu(\bar\psi(x)\gamma_\mu\gamma_5\psi(x))=0~$  respectively. However, in QFT, the composite  operators themselves, being products of operators at the same spacetime point, are ill-defined due to the inherent short distance singularity, thereby making the above equations  ambiguous.   It becomes necessary to work with suitably regularized quantum operators. Once the regularization is removed after mathematical operations are performed, the conservation equations can get modified giving rise to anomalies. In  massless QED, both of the above conservation laws can not be maintained. One chooses to preserve  charge  conservation, $\partial^\mu J^V_\mu=0$, thus breaking the axial current conservation law, $\partial^\mu J^A_\mu= \frac{e^2}{4\pi^2}{\mathbf E}.{\mathbf B}$,  RHS being the divergence anomaly in axial current, in terms of electric and magnetic fields. Generically, anomalies act in two distinct way: in external currents (not coupled to gauge interactions) they can induce physical effects such as $\pi_0$ decaying into two photons via axial anomaly. On the other hand consistency of chiral gauge theories demand vanishing of total anomaly (t'Hooft's anomaly cancellation condition) thereby constraining viable models, to wit the Standard Model or Superstring theory (via Green–Schwarz mechanism) \cite{tr}.

Another crucial aspect of QFT is  current algebra (see for example \cite{gold}), where certain equal time commutation relations among  current
density operators define an infinite-dimensional Lie algebra. Originally proposed by Gell-Mann to describe
 strongly interacting hadron physics, it led to  Adler–Weisberger
formula and other important physical results and initiated theoretical developments such as Sugawara model, light
cone currents, Virasoro algebra, the mathematical
theory of affine Kac–Moody algebras, and nonrelativistic current algebra in quantum and statistical physics. Specific  applications include  conformally invariant
field theory, vertex operator algebras, exactly
solvable lattice and continuum models in statistical
physics, exotic particle statistics and q-commutation relations, hydrodynamics and quantized vortex
motion.

Returning to the present context,  anomalies  can modify the current algebra, {\it{i.e.}}  through the introduction of Schwinger terms \cite{sj,sch}, computed along similar lines as divergence anomaly. As proved by Adler \cite{bran} and studied by others \cite{mitra}, the divergence anomaly and Schwinger terms (or commutator anomalies) are complimentary effects - presence of one type necessitates existence of the other (in fact om 1+1-dimensions the Schwinger terms can uniquely yield the divergence anomaly \cite{mitra}) - and the anomalous extensions have to obey the Adler \cite{bran} consistency condition. This will play a major role in our work.

In recent years divergence anomaly has generated a huge amount of interest in an unexpected scenario -   hydrodynamic regime of a non-relativistic  classical field theory. Effect of anomalous current algebra on  Raman scattering  in Mott insulators was studied in \cite{wig1}. Son and Spivak \cite{spi} have shown that large classical negative magnetoresistance of  Weyl metals  is connected to the triangle anomaly in  classical regime where   mean free path of the  electron is short compared to the magnetic length. Further works in related areas are \cite{sur,nair}. In \cite{sur} quantum anomalies for global currents in hydrodynamic limit and its novel consequences were revealed. Gauge anomalies present in in hydrodynamics were also studied in Hamiltonian formulation in \cite{nair}. After reporting our results \cite{arx}  we came across the paper  \cite{wig} that also discusses  divergence anomaly in classical fluid.

All the recent excitement in condensed matter  phenomena exploits the divergence anomaly. Quite surprisingly, in the literature the (divergence) anomaly is simply assumed to exist in an ad hoc way and  there are no attempts to {\it{derive}} it from first principles. Furthermore there are no mentions of Schwinger terms in these recent works.
\vskip .5cm
{\bf{Significance and new results of our work:}}  In this Letter we will shed some light on the two untouched problems mentioned above in the following format:\\
{\bf{(I)}} first principle derivation of anomalous current algebra ($\sim$ Schwinger terms) in classical fluid dynamics;  \\
{{\bf{(II)}}  derivation of the classic chiral anomaly form $e^2{\mathbf E}.{\mathbf B}$  in helicity conservation equation from the contribution of the above mentioned Schwinger terms;}\\
{\bf{(III)}} establishing analogy between the generalized velocity field defined here and the  well known "anomalous velocity", appearing in condensed matter physics \cite{hall}.\\
{\bf{(IV)}} demonstration of  Schwinger terms and divergence anomaly satisfying  Adler consistency condition \cite{bran}. \\
{{\bf{(V)}} Construction of  Casimir operators (in weak field approximation) in the anomalous system via  Darboux prescription.\\
The procedure adopted here is systematic and unambiguous but, at the same time,  we emphasize that application of the above   approach and ensuing results in the context of  fluid anomaly  is completely new.} \\
 {\bf{(VI)}} Topical interest of the work will be mentioned presently.

To simplify the nomenclature we refer to any non-canonical or anomalous term in algebra among dynamical variables as Schwinger term. In our scheme, anomalous current algebra plays the primary role  and divergence anomaly appears naturally as a derived quantity. We  explicitly show that the anomalous fluid model can be constructed in a natural way in  Hamiltonian formalism in Eulerian approach.  The quantum input comes from basic generalized Poisson bracket structure satisfied by  the discrete (Lagrangian) fluid particle coordinates. This phase space characterizes the  semi-classical  electron dynamics in a magnetic Bloch band, in presence of a   periodic potential with external  magnetic field and Berry curvature \cite{niu,hor,ber, ch}. Berry curvature and induced magnetic field in momentum space is responsible for the "anomalous velocity" in quantum Hall effect \cite{hall}. Nonzero fluxes of the Berry curvature in  electron Fermi surface plays an important role in transport phenomena.

To further emphasize the prospects of the present work, we note that the history of electron hydrodynamics in condensed matter, {\it{i.e.}} situations where electron flow is influenced by  hydrodynamic laws instead  being fully Ohmic is nearly fifty years old \cite{ssg}. In normal circumstances,   electrons in metals behave as a nearly-free Fermi gas since the effective mean free path for electron-electron collision is quite large which allows   impurities and lattice
thermal vibrations (phonons) to destroy a  collective viscous fluid-like motion  of electrons. However, in recent years the hydrodynamic regime has been achieved   in extremely pure   high quality
electronic materials - especially graphene \cite{graf}, layered materials with very high electrical conductivity such as metallic delafossites  $PdCoO_2  ,~ PtCoO_2$ \cite{met} among others.

The all important ingredient in our analysis is the well known map (constitutive relation) that expresses the continuous Euler fluid variables in terms of the discrete Lagrangian particle coordinates. Through this map the fluid field algebra inherits the anomalous features (Schwinger terms) from the quantum corrected Poisson brackets of Lagrangian  coordinates and subsequently divergence anomalies follow from Hamiltonian equations of motion. The flowchart of our scheme follows: \\
generalized phase space algebra with Berry curvature corrections \cite{niu,hor, ch} $\rightarrow$ (via constitutive relations) extended fluid variable algebra $\rightarrow$ extended fluid equations with divergence anomaly and Schwinger terms $\rightarrow$ consistency condition connecting divergence anomaly and Schwinger terms.

\vskip .5cm
 {\bf{(I) Derivation of anomalous fluid algebra  (Schwinger terms):}} We begin with the  Berry phase corrected ($\sim$ anomalous or noncommutative) phase space algebra of the D.O.F. $X_{j}({\bf x}),~P_j({\bf x})=M\dot{X}_{j}({\bf x})$
 ($M$ being the point particle mass), to be identified subsequently with the discrete individual particle (Lagrangian) phase space coordinates \cite{niu,hor,ber,ch},
\begin{equation*}
\lbrace X_{i}({\bf x}), X_{j}({\bf x'})\rbrace=- \frac{1}{\rho_0}\epsilon_{ijk}\mathcal{F}_{k}~\delta({\bf x}-{\bf x'});~~
\lbrace X_{i}({\bf x}), P_{j}({\bf x'})\rbrace= \frac{M(\de_{ij}+e B_{i}\Omega_{j})}{\rho_0\mathcal{A}}\delta({\bf x}-{\bf x'}); 
\end{equation*}
\begin{equation}
    \label{posit}
    \lbrace P_{i}({\bf x}), P_{j}({\bf x'})\rbrace= e\frac{\epsilon_{ijk} M^2 B_{k}}{\rho_0\mathcal{A}}\delta({\bf x}-{\bf x'})
\end{equation}  
where 
$$\mathcal{F}_{i}({\bf x,\bf k})=\frac{\Omega_{i}}{1+e{\bf{B}}({\bf x}).{\bf{\Omega}(\bf k)} },~~	\mathcal{A}({\bf x,\bf k})=1+e{\bf{B}}({\bf x}).{\bf{\Omega}(\bf k)}.$$
In the above Lagrangian degrees of freedom, coordinate and velocity  $X_{(n)i},P_{(n)i}$ respectively, the discrete particle index $n$ is replaced by ${\bf x}$ in the continuum limit. Furthermore, $\rho_0$ is a dimensionful  parameter, $e$ is the electronic charge, ${\bf{B}}$ the external magnetic field and ${\bf{\Omega}}({\bf {k}})=\nabla_{{\bf {k}}}\times {\bf {A}}({\bf {k}})$ the Berry curvature  in momentum space.  For ${\bf{\Omega}}=0, {\bf{B}}=0$ one recovers the free canonical Poisson brackets. ${\bf{\Omega}}$ appears as a result of electron motion in a periodic lattice potential.  These non-relativistic brackets are at equal time.

It is worthwhile to point out that in general $\bf P $ and $\bf k$ (the crystal momentum) are in fact related. However, as has been considered in \cite{new}, in a toy model approach $\bf \Omega $ can be treated as a constant and furthermore explicit calculations with realistic models have shown \cite{new} that it is of the order of $(lattice~constant)^2$. Throughout our subsequent work and mainly for computational convenience, we will restrict $\bf \Omega $ to be independent of dynamical variables. At the end we will comment on some other non-trivial possibilities. 

 Since $M$ is a constant parameter, we will drop it to express formally the above brackets (\ref{posit}) using ${{X}_{i}(\bf x), \dot{X}_{i}(\bf x)}$. This will also help in streamlining the notation because the constitutive relations connecting   the Euler  variables density $\rho({\bf r})$ and velocity fields $v_i({\bf r}) $ are  directly connected to them (see for example \cite{jac}),
 \begin{equation}
  \rho({\bf r})=\rho_0\int dx~\delta(X(x)-{\bf{r}}),~~
  v_i({\bf r})=\frac{\int dx ~\dot{X_i}(x)\delta (X(x)-{\bf{r}})}{\int dx ~ \delta(X(x)-{\bf{r}})}.
 \label{ncc13}
 \end{equation}
 Note that $j_i=\rho v_i$ constitute the momentum density of the fluid. Even though we are not directly dealing with relativistic fluid dynamics and our system is without dissipative effects, still it is worthwhile to mention that this definition pertains to the  Landau frame where $j_i(\vec r,t)$ refers to the values at a fixed spacetime position $\vec r,t$. The Landau frame is chosen in the direction of the total energy where the directions of
the eigenvector of the energy-momentum tensor and the
conserved current match. This is generally true in  non-relativistic hydrodynamic 
flow which can be defined as
a local 
flux of particles.

  Using (\ref{posit}) it is straightforward to compute the anomalous fluid brackets,
\begin{equation}
\label{roro}
\{\rho({\bf{r}}), \rho({\bf{r'}})\}= \epsilon_{ijk}\pa_{i}^{{\bf{r}}}(\rho({\bf{r}}) \mathcal{F}_{k}(\bf r)\pa_{j}^{{\bf{r}}}\de({\bf{r}}-{\bf{r}'})
\end{equation}

\begin{equation}
	\label{rov}
	\{\rho({\bf{r}}), v_{i}({\bf{r'}})\}=\frac{\pa_{i}^{{\bf{r}}}\de({\bf{r}}-{\bf{r}'})}{\mathcal{A}({\bf{r'}})}+ e B_{j}({\bf{r'}})\mathcal{F}_{i}({\bf{r'}}))\pa_{j}^{{\bf{r}}}\de({\bf{r}}-{\bf{r}'})+\epsilon_{ljk}\mathcal{F}_{l}({\bf{r'}})\pa_{j}^{{\bf{r'}}}v_{i} \pa_{k}^{{\bf{r}}}\de({\bf{r}}-{\bf{r}'})
	\end{equation}
	
\begin{equation}
\nonumber
\{v_{i}({\bf{r}}), v_{j}({\bf{r'}})\}= \lbrace\frac{\pa_{j}v_{i}-\pa_{i}v_{j}}{\rho\mathcal{A}({\bf{r}})} +e\epsilon_{ijk}\frac{B_{k}({\bf{r}})}{\rho\mathcal{A}({\bf{r}})}
\end{equation}	

\begin{equation}
\nonumber
-\frac{e B_{l}({\bf{r}})}{\rho}(\mathcal{F}_{i}({\bf{r}})\pa_{l}v_{j}-\mathcal{F}_{j}({\bf{r}})\pa_{l}v_{i})-2\epsilon_{lmn}\frac{v_{i}}{\rho^{2}}\pa_{m}v_{j}\pa_{l}(\mathcal{F}_{n}({\bf{r}})\rho)
\end{equation}
\begin{equation}
\nonumber
-\epsilon_{lmn}\frac{1}{\rho}\pa_{n}v_{i}\pa_{m}v_{j}\mathcal{F}_{l}({\bf{r}}) -2\epsilon_{lmn}\frac{\pa_{m}\rho}{\rho^{2}}\pa_{n}(v_{i}v_{j})\mathcal{F}_{l}({\bf{r}}) \rbrace\de\brr
\end{equation}
	\begin{equation}
	\label{vv}
	 +2\epsilon_{lmn}\frac{\pa_{m}^{{\bf{r}}}\delta \brr}{\rho({\bf{r}})}\pa_{n}^{{\bf{r}}}(v_{i}v_{j})\mathcal{F}_{l}({\bf{r}}).
	\end{equation}
	
	We refer to the non-canonical ${\bf{\Omega}}$-dependent terms as Schwinger terms. It is very important to stress that, although Berry curvature  ${\bf{\Omega}}$ behaves as an effective magnetic field, its effect is distinct from the effect of the true external magnetic field ${\bf{B}}$. In this connection, note that this  anomalous fluid algebra is different from the general structure derived from the results in \cite{mor}, where ${\bf{B}}, {\bf{E}}$ are non-vanishing but  ${\bf{\Omega}}=0$.   However, the algebra matches with the brackets used in \cite{nair} for ${\bf{\Omega}}=0$. We emphasize that even if the (canonical) vorticity  $\omega_{ij}=\pa_{i}v_{j}-\pa_{j}v_{i}$ vanishes, effective vorticity can reappear anomalously, as  seen in (\ref{vv}). Similar types of extended fluid brackets in different context have appeared in \cite{sg1}. The canonical  brackets for $\rho_c,v_{(c)i}$  are recovered for ${\bf B}={\bf \Omega }=0$,
\begin{equation}
\nonumber
\{\rho_c({\bf{r}}), \rho_c({\bf{r'}})\}=0,~~\{\rho_c({\bf{r}}), v_{(c)i}({\bf{r'}})\}=\pa_{i}^{{\bf{r}}}\de({\bf{r}}-{\bf{r}'}),~~
	      \end{equation}
	     \begin{equation}
   	     \{v_{(c)i}({\bf{r}}), v_{(c)j}({\bf{r'}})\}= -\frac{(\pa_{i}v_{(c)j}-\pa_{j}v_{(c)i})}{\rho_c}\de({\bf{r}}-{\bf{r}'}).
   	     \label{free}
   	     \end{equation}

The anomalous fluid algebra (\ref{roro}), (\ref{rov}), (\ref{vv}) constitute first part of our work.	

 It is important to note that $j_i=\rho v_i$ being the momentum density, should act as  translation generator, in absence of external field $B$. This is indeed the case for  arbitrary functions $\alpha(\rho),~A_{l}(v_{i})$ as shown below (computational details are provided in Supplementary Material); 
 \begin{equation}
	\label{rojifinal}
    \lbrace \alpha(\rho(x)) ,\int d^3y~ \rho v_{l} \rbrace= \frac{d\alpha}{d\rho}\pa_{l}\rho = \pa_{l}\alpha(\rho(x)) 
\end{equation} 
  \begin{equation}
	    \label{vjifinal}
	     \lbrace A_{l}(v_{i}(x)) , \int d^3y~ \rho v_{j} \rbrace = \frac{d A_{l}}{d v_{i}}  \pa_{j}v_{i}  = \pa_{j} A_{l}(y) .
	\end{equation}

\vskip .3cm
{\bf{(II) Conservation law and helicity  anomaly:}} We quickly recapitulate derivation of canonical fluid equations of motion.  Together with the Hamiltonian for a barotropic fluid (with the pressure $P=\rho (dV)/(d\rho )-V$ depending only on density $\rho$)
\begin{equation}
	     \label{ham}
	     \mathcal{H}_0=\int d{\bf x}(\frac{1}{2}\rho v^2 +V(\rho)) 
	 \end{equation}
and the brackets (\ref{free}), the canonical continuity and Euler equations are obtained as
\begin{equation}
	     \label{hamo}
\dot \rho({\bf x})=\{\rho({\bf x}), \mathcal{H}_0\}= -\nabla(\rho {\bf v}),~~
\dot {{\bf v}}({\bf x})=\{{\bf v}({\bf x}), \mathcal{H}_0\}=- ({\bf{v}}.{\bf{\nabla}}){\bf{v}}-\frac{{\bf{\nabla}}P}{\rho}.
\end{equation}
  The fluid Hamiltonian in external electromagnetic field is  given by
	 \begin{equation}
	     \label{ham1}
	     \mathcal{H}=\int d{\bf x}~( \frac{1}{2}\rho v^2 +V(\rho)-e\rho\Phi )
	 \end{equation}
	  where the electric field is  ${\bf{E}}= -\nabla \Phi$ in a time independent scenario. The effect of magnetic field ${\bf B}$ has already been taken into account through the extended symplectic  structure (\ref{roro}), (\ref{rov}), (\ref{vv}). The continuity equation gets modified to
	\begin{equation}
	\label{con}
	\dot{\rho}+{\bf{\nabla}}.{\bf J}^{an}=e\rho {\bf{\mathcal{F}}}.({\bf{\nabla}}\times {\bf E}) 
	\end{equation}
where the anomalous current ${\bf J}^{an}$ is
\begin{equation}
	\label{J}
{\bf J}^{an}=(\frac{\rho{\bf{v}}}{\mathcal{A}})+ e \rho({\bf{\mathcal{F}}}.{\bf{v}}){\bf{B}}+e \rho ({\bf E}\times {\bf{\mathcal{F}}})+\mathcal{F}\times \nabla P .
\end{equation}
However, in the present time independent case the conservation law $ \dot{\rho}+{\bf{\nabla}}.{\bf J}^{an}=0$ survives, albeit with an anomalous current, since  now  Maxwell's equation yields ${\bf{\nabla}}\times {\bf E}=-(\partial {\bf{B}})/\partial t=0.$ 

The anomalous Euler equation is derived as
\begin{equation}
\nonumber
\dot{{\bf{v}}}+ \frac{({\bf{v}}.{\bf{\nabla}}){\bf{v}}}{\mathcal{A}}=-\frac{{\bf{\nabla}}P}{\rho\mathcal{A}} +e\frac{\rho {\bf{v}}\times{\bf{B}}}{\rho\mathcal{A}}-e \frac{{\bf{B}.{\bf{\nabla}}P}}{\rho}{\bf{\mathcal{F}}}-e({\bf{v}}.{\bf{\mathcal{F}}})({\bf{B}}.{\bf{\nabla}}){\bf{v}}+\lbrace (\frac{\nabla P}{\rho}\times \mathcal{F}).\nabla
\end{equation}	
	\begin{equation}
	\label{eul}
	-\frac{1}{\rho}\nabla v^2 .(\nabla \times  {\bf({\mathcal{F}\rho)}})+ 2 v^2 ({\bf{\mathcal{F}}}\times \frac{\nabla \rho}{\rho}).\nabla - {\bf{\mathcal{F}}}. (\frac{\nabla \rho}{\rho}\times \nabla v^2))\rbrace{\bf{v}}
	-e\lbrace \frac{{\bf{E}}}{\mathcal{A}}+ e ({\bf{E}}.{\bf{B}}){\bf{\mathcal{F}}}-({\bf{E}}\times{\bf{\mathcal{F}}}). {\bf{\nabla}}{\bf{v}}\rbrace
	\end{equation}
The most interesting term in the RHS of (\ref{eul})	is obviously $e^2  ({\bf{E}}.{\bf{B}}){\bf{\mathcal{F}}}$ since it has the classic chiral anomaly form.

An important property of fluid dynamics is the vorticity ${\bf \omega =\nabla \times v}$ and the related pseudo-scalar quantity helicity $ \Sigma =\int d{\bf x}~h=\int d{\bf x}~{\bf v.\omega }$. In Newtonian fluid helicity is conserved and  satisfies a local conservation law. In the present scenario, the modified time-evolution  for helicity can be expressed as
\begin{equation}
	\label{h1}
\dot h=\dot {\bf v}.\omega +{\bf v}.\dot \omega =-\nabla .({\bf v}\times \dot {\bf v}) +2\dot {\bf v}.\omega .	
	\end{equation}
Ignoring the divergence term that will drop off upon space integration and using (\ref{eul}) we find 	
	\begin{equation}
	\nonumber
\dot h/2=-e^2({\bf E}.{\bf B})(\omega .\mathcal{F})-e\frac{(\omega .{\bf E})}{\mathcal{A}}	-e\frac{({\bf B}.\nabla P)}{\rho}(\omega .\mathcal{F})-e\frac{\omega .({\bf v}\times {\bf B})}{\mathcal{A}}-e({\bf v}.\mathcal{F})[\omega .({\bf B}.\nabla ){\bf v}] 
	\end{equation}
	\begin{equation}
	    \label{hh23}
	     +\Psi h +\omega .[\mathcal{F}.({\bf C}\times \nabla )]{\bf v}
	\end{equation}
where
$$\Psi = -\nabla v^2 .(\nabla \times \mathcal{F})+\nabla v^2 .(\mathcal{F}\times \frac{\nabla \rho}{\rho }); ~~{\bf C}=-\frac{\nabla P}{\rho}+\nabla v^2-e{\bf E}.$$
	
	Let us  consider some simple and restrictive cases; (a) Only the first three terms in the RHS are of $O(v)$ with  rest of the terms being of higher order in $v$. Hence in a low energy regime these terms can be ignored.\\
	(b) We can consider pressure $P$ to be absent.\\
(c) Rewriting $\omega .{\bf E}=(\nabla \times {\bf v}).{\bf E}=\nabla.({\bf v}\times {\bf E})+{\bf v}. (\nabla \times {\bf E})$ we can consider time independent external ${\bf E}, {\bf B}$ configurations so that 	$\nabla \times {\bf E}=0$ and hence the second term also will not contribute. Thus we are left with the chiral anomaly term
\begin{equation}
\label{anomaly}
\dot \Sigma=-2e^2({\bf E}.{\bf B})\int d{\bf {X }}(\omega .\mathcal{F}).
\end{equation}
This is the cherished form of anomaly in Eulerian fluid (in low energy limit) and constitute another major result.

\vskip .3cm
{\bf{(III) Analogy between Generalized and "anomalous" velocity \cite {hall,niu,hor, ch}:}}
Let us rewrite  ${\bf {J}}^{an}$ from (\ref{J}) in a more suggestive form as 
\begin{equation}
    \label{cd}
  {\bf {J}}^{an}=(\frac{\rho}{1+e{\bf{B}}.{\bf{\Omega}}})(
  {\bf{v}}+ e \rho({\bf{\Omega}}.{\bf{v}}){\bf{B}}+e \rho ({\bf E}\times {\bf{\Omega}})+(\frac{{\bf {\Omega}}}{1+e{\bf{B}}.{\bf{\Omega}}})\times \nabla P .
  \end{equation}  
  Keeping terms of $O(e)$ only we write ${\bf {J}}^{an}\approx \rho {\bf {v}}^{gen}$ where, ignoring the pressure term, the generalized velocity is
  \begin{equation}
    \label{cd1}
  {\bf {v}}^{gen}=   (1-e{\bf{B}}.{\bf{\Omega}}){\bf{v}}+ e \rho({\bf{\Omega}}.{\bf{v}}){\bf{B}}+e \rho ({\bf E}\times {\bf{\Omega}}).
    \end{equation}

  On the other hand, following \cite {hall,niu,hor, ch}, from the Bloch electron dynamics in a magnetic band with $\epsilon_n({\bf {k}})$ being the $n$-th band energy, 
  \begin{equation}
    \label{cdd}
    {\bf {\dot r}}=\frac{\partial \epsilon_n({\bf {k}})}{\partial{\bf {k}} } -{\bf {\dot k}}\times {\bf {\Omega}},~~{\bf {\dot k}}=-e{\bf {E}}-e{\bf {\dot r}}\times {\bf {B}}
   \end{equation} 
   we obtain 
    \begin{equation}
    \label{cd2}
   {\bf {\dot r}}=\frac{\partial \epsilon_n({\bf {k}})}{\partial{\bf {k}} }+e {\bf {E}}\times {\bf {\Omega}}+e({\bf {\dot r}}\times {\bf {B}})\times {\bf {\Omega}},
   \end{equation} 
   where ${\bf {\dot r}}$ in (\ref{cd2}) is referred as the "anomalous velocity" \cite {hall,niu,hor, ch}. Note the the $e$-dependent terms in (\ref{cd1}) and (\ref{cd2}) are identical. This matching clearly shows that the anomaly inherited by the ideal classical fluid (from the Bloch band electron dynamics) has a deeper significance. This will be further strengthened  in the next section where we prove consistency of the full anomalous structure developed in this Letter.

   \vskip .3cm
 {\bf{(IV) Consistency condition involving divergence anomaly and Schwinger term:}} In this section we restrict to constant external fields ${\mathbf {E}},{\mathbf {B}}, {\mathbf {\Omega}} $ but the results can be easily extended to non-constant external fields. Let us rewrite the equations (\ref{con}), (\ref{roro}) and (\ref{rov}) schematically as
 \begin{equation}
    \label{c1}
   	\dot{\rho}+{\bf{\nabla}}.{\bf J}^{an}=0,
   \end{equation} 
\begin{equation}
    \label{c2}	
   	   	\{\rho({\bf{r}}), \rho({\bf{r'}})\}=S({\bf{r}},{\bf{r'}});~\{\rho({\bf{r}}), J^{an}_i({\bf{r'}})\}=S_i({\bf{r},{\bf{r'}}}).
\end{equation}

Taking time derivative of both sides of the generic bracket  $\{\rho, \rho\}=S $ equation in (\ref{c2}) we get 
\begin{equation}
    \label{cc1}
  \partial_0  \{\rho({\bf{r}}), \rho({\bf{r'}})\} =\{\partial_0\rho({\bf{r}}), \rho({\bf{r'}})\} + \{\rho({\bf{r}}), \partial_0\rho({\bf{r'}})\}=\partial_0S({\bf{r}},{\bf{r'}}).
\end{equation}
Using the bracket (\ref{roro}),  the RHS of (\ref{cc1}) is given by 
\begin{equation}
    \label{ss}
\partial_0S({\bf{r}},{\bf{r'}})=\partial_j^r( \epsilon_{ijk}\mathcal{F}_{k}\partial_i(\dot \rho )\de({\bf{r}}-{\bf{r}'}))=-\partial_j^r( \epsilon_{ijk}\mathcal{F}_{k}\partial_i(\partial_l J^{an}_l)\de({\bf{r}}-{\bf{r}'}))
\end{equation}
where $J^{an}_l$ is given by (\ref{J}). In LHS of (\ref{cc1}) we have 
\begin{equation}
    \label{ss1}
\{\partial_0\rho({\bf{r}}), \rho({\bf{r'}})\} + \{\rho({\bf{r}}), \partial_0\rho({\bf{r'}})\}=-\{\partial_i J^{an}_i({\bf{r}}), \rho({\bf{r'}})\} - \{\rho({\bf{r}}), \partial_i J^{an}_i({\bf{r'}})\}.
\end{equation}
Again substituting $J^{an}_i$ from (\ref{J}) and computing each bracket from the full bracket structure in Supplemental Material (1,2,3), after a long algebra we recover (\ref{ss}), thereby ensuring that the consistency condition is satisfied. 
 \vskip .3cm
   {\bf{(V) Darboux transformation and Casimir Operators:}} As we have shown the anomalous fluid model is an extension of the canonical Hamiltonian fluid and since the latter system has two Casimir operators, total $\rho $-charge $h$-and helicity, (arising out of the 
relabeling symmetry  in the Euler formulation for fluids), it becomes imperative to construct extensions of those Casimirs in the anomalous fluid model. In simple systems  generally  ever in complicated systems one can fall back to Darboux theorem that guarantees, at least locally,  that one can construct combinations of non-canonical variables that behave canonically.   Instead of working directly with the continuous fluid variables, it is easier to construct the Darboux map in the discrete noncommutative variables (\ref{posit}) where the following combinations are canonical, up to $O(e\bf{E},e\bf{B}, {\bf{\Omega }})$  for simplicity:
\begin{equation}
\label{dar}   X_i= q_{i} + \frac{1}{2M}\epsilon_{ijk}{p_{j}}\Omega_{k},~~
P_i= p_{i}+\frac{eM}{2}\epsilon_{ilk}q_{l}B_{k},
\end{equation}
where $\{q_i,q_j\}=\{p_i,p_j\}=0, \{q_i,p_j\}=\delta_{ij}$.

Utilising the basic definition of fluid variables in terms of discrete degrees of freedom (\ref{ncc13}) it is straightforward to generate the  combinations of anomalous variables that behave canonically (subject to the approximation mentioned above):

\begin{equation}
\label{jd}\rho_c({\bf{r}})=\rho ({\bf{r}}) - \frac{1}{2}\epsilon_{ijk}\Omega_{k}\pa_{i}j_{j}({\bf{r}})
\end{equation}
\begin{equation}
\label{jd1}
j_{(c)i}({\bf{r}})= j_{i}({\bf{r}}) -\rho_{0}\int dx [ \frac{1}{2}\epsilon_{ljk}\Omega_{k}\dot{X_{i}}\dot{X_{j}}\pa_{l}^{{\bf{r}}}\delta({\bf{X(x)-r}}) -\frac{e}{2}\epsilon_{ilk}X_{l}B_{k}\delta({\bf{X(x)-r}}) ]
\end{equation}

It is now straightforward to derive the map between $\rho_c,v_{(i)c}$ and its anomalous counterpart $\rho,v_{(i)}$ and subsequently construct the Casimir operators in anomalous phase space. Infact in (\ref{jd}) we already have one of the cherished Casimirs and we have not shown the explicit form of the other Casimir, i.e. helicity. A peculiarity of (\ref{jd1}) is that the RHS is not closing in terms of $\rho,j_i$ and higher moments are coming in to play due to the essential non-linearity in the model. 

{\bf{(VI)  Non-trivial forms of $\bf\Omega $:}} As we have mentioned earlier, besides the constant form  used here, there are non-trivial forms of $\bf\Omega $ that are of topical interest in Condensed Matter Physics, such as\\
(i) In semiclassical framework of  Anomalous Hall Effect \cite{hall, anhal} in metallic ferromagnets   induces an anomalous velocity contribution
to Bloch wavepacket group velocity, generated by  momentum-space Berry curvatures. In this case, the anomalous Hall conductivity $\sigma_{ij} =-\epsilon_{ijl}\frac{e^2}{\hbar }\Omega ^l$ consists of the Berry phase $\bf\Omega $, a linear combination of reciprocal lattice vectors $\bf G$. In this case our formalism can be carried through by simply substituting the explicit expression for  $\bf\Omega $ in density ($\rho $) and current ($j_i$) operators.\\
(ii) Another well known example of non-trivial Berry phase is in  Weyl  semimetal \cite{weyl};  $\Omega_i=\pm p_i/(2|p|^3)$. 
A Weyl semimetal is a 3D crystal whose low energy excitations are Weyl fermions. The Berry curvature  monopoles are located at Weyl points in the Brillouin zone. Unfortunately due a  naive adaptation of our formalism for this particular case  might be computationally problematic due to the presence of   $v_i$ in  RHS of the anomalous brackets in (\ref{roro}, \ref{rov},\ref{vv}).

\vskip .3cm
{{\bf{Discussion:}} In this paper, we have developed an extended classical fluid model incorporated with Berry phase effects that generates  a divergence anomaly $\sim ~ e^2{\bf E}.{\bf B}$ (having the form of Adler-Bell-Jackiw chiral anomaly) in helicity conservation equation.  We have shown a direct analogy between the generalized fluid velocity field  defined here and the well known "anomalous velocity" appearing in quantum Hall effect in condensed matter physics. Overall validity of the entire anomalous structure is demonstrated through the satisfaction of Adler consistency condition. We have developed a systematic program to construct the Casimir operators for the anomalous fluid model. Lastly we have discussed applicability of our scheme in some specific Berry curvature structures that are of interest in Condensed Matter Physics.

Our approach is semiclassical where we have constructed the anomalous  fluid  bracket structure based  on Poisson brackets augmented with (quantum mechanical) Berry phase effect.  The latter is applicable for electrons moving in magnetic Bloch bands. Hence our anomalous fluid model can have relevance in hydrodynamic equations describing electron gas models  subject to spin-orbit-like interactions in condensed matter systems, such as graphene \cite{gr}. In condensed matter systems in a semi-classical framework,  transport is studied through Boltzmann equation involving the distribution function $f({\mathbf {x}},{\mathbf {p}}, t) $ and the density function appearing here is $\rho({\mathbf {x}},t)=\int mf~d{\mathbf {v}}$. Thus the anomalous equations revealed here will alter the Boltzmann equation. Electron transport in Bloch bands in hydrodynamic limit are given by Eulerian fluid equations which should be modified appropriately. An interesting recent case in point is the work \cite{song} where transport of  collective excitations, named as chiral Berry plasmons, in hydrodynamic limit, is studied in generic  interacting metallic systems with nonzero Berry flux. In \cite{zak},   chiral liquids,  consisting of right-left asymmetric massless fermions is considered where  electromagnetic current in presence of  external magnetic
field will carry a chiral anomaly.

Previously, $O(\hbar ^2)$ corrections were introduced in classical fluid equations \cite{gard} from  a moment expansion of the Wigner-Boltzmann equation. Interestingly in the present work, the Berry curvature plays an essential role in inducing $O(\hbar )$ correction. This can be seen by comparing a classical model Lagrangian
	\begin{equation}
	\label{d1}
	L=\frac{1}{2}m {\bf v}^{2} - e \Phi + e {\bf A}.{\bf v}
	\end{equation}
	with $e$ dimensionless and $[{\bf B}]=\frac{M}{t},~[{\bf E}]=\frac{ML}{t^2} $
		and the definitions of  Lorentz force and electromagnetic fields respectively 
$$ {{\bf F}}=\nu({\bf E}+ {\bf v}\times{\bf B}),~~~ {\bf E}=-\nabla\Phi -\frac{\pa {\bf A}}{\pa t},~~~{\bf B}=\nabla\times {\bf A}$$ with a quantum Lagrangian
\begin{equation}
    \label{lagq}
  L=  \hbar k \dot{{\bf r}}-e \dot{{\bf r}}\times {\bf A} + \hbar \dot{k}{\bf A_{\beta}} + e \Phi - W
\end{equation}
$[\hbar k]= \frac{ML}{t}$,   $[A_{B}]=L$,   $[\mathcal{F}]=\frac{L}{M \frac{L}{t}}= \frac{t}{M}$ with 
$[{\bf B}.\Omega]$ being dimensionless. In the above $A_{B}$ is the Berry potential and $\Omega =\nabla_{p}\times A_{B}$ the Berry curvature. Coming back to the fluid variables, its dimensions are
$[\rho ]= \frac{M}{L^3} $, $[{\bf v}]=\frac{L}{t}$.

 Abanov and Wiegmann \cite{wig}  have also shown the presence of chiral anomaly in a generalized helicity conservation equation. However the framework is entirely different from ours and this is reflected in the difference between the explicit  anomaly equations. In \cite{wig} the anomaly appears to be induced as a non-inertial effect. On the other hand we have followed throughout a systematic Hamiltonian approach from first principles, starting from a semi-classical Poisson algebra with Berry phase corrections that induces a generalized (anomalous) fluid algebra. Subsequently the Hamiltonian equations of motion yields the generalized continuity and Euler equations leading to the anomaly. Furthermore, current algebra  and Schwinger terms do not play any role in \cite{wig}.

\vskip .3cm
{\bf Acknowledgement:} S.G. would like to thank Professor Peter Horvathy for correspondence. We thank the Referees for constructive comments.

		\end{document}


\title{{\bf{Supplementary file for ``Divergence Anomaly and Schwinger Terms:\\
Towards a Consistent Theory of  Anomalous  Classical Fluid"
 }}}
	\author{  {\bf {\normalsize Arpan Krishna Mitra$^1$}$
			$\thanks{E-mail: arpankrishnamitra@hri.res.in}}$~$
		{\bf {\normalsize Subir Ghosh$^2$}$
			$\thanks{E-mail: subir\_ ghosh2@rediffmail.com}},$~$
		\\{\normalsize $^1$Harish Chandra Research Institute, HBNI,}
		\\{\normalsize Chhatnag Rd, Jhusi, Uttar Pradesh 211019}
		\\\\
		{\normalsize $^2$Indian Statistical Institute}
		\\{\normalsize  203, Barrackpore Trunk Road, Kolkata 700108, India}
		\\[0.3cm]
	}
	\date{}
	\maketitle

The following important set of brackets is derived from the definition $j_i=\rho v_i$,	
\begin{equation}
\label{roro}
\{\rho({\bf{r}}), \rho({\bf{r'}})\}= \epsilon_{ijk}\partial_{i}^{{\bf{r}}}(\rho({\bf{r}}) \mathcal{F}_{k})_{r}\partial_{j}^{{\bf{r}}}\de({\bf{r}}-{\bf{r}'})
\end{equation}
\begin{equation}
\label{roj}
\{\rho({\bf{r}}), j_{i}({\bf{r'}})\}=[\frac{\rho({\bf{r'}})({\bf{r'}})\partial_{i}^{{\bf{r'}}}\de({\bf{r}}-{\bf{r}'})}{\mathcal{A}}+ e B_{j}\mathcal{F}_{i}({\bf{r}})\rho({\bf{r'}})\partial_{j}^{{\bf{r}}}\de({\bf{r}}-{\bf{r}'})+\epsilon_{ljk}\partial_{j}^{{\bf{r}}}(j_{i}({\bf{r}}) \mathcal{F}_{l})_{r}\partial_{k}^{{\bf{r}}}\de({\bf{r}}-{\bf{r}'})]
\end{equation}

\begin{equation}
\nonumber
\{j_{i}({\bf{r}}), j_{j}({\bf{r'}})\}=\frac{\nu\epsilon_{ijk} B_{k}\rho}{\mathcal{A}} \de({\bf{r}}-{\bf{r}'})-(\frac{\de_{li}}{\mathcal{A}}+\nu B^{l}\mathcal{F}_{i})_{{\bf{r}}}j_{j}({\bf{r}})\partial_{l}^{{\bf{r'}}}\de\brr  -(\frac{\de_{lj}}{\mathcal{A}}+\nu B^{l}\mathcal{F}_{j})_{{\bf{r'}}}j_{i}({\bf{r'}})\partial_{l}^{{\bf{r'}}}\de\brr
\end{equation}
\begin{equation}
\label{jj}
+\epsilon_{lmn}\partial_{l}^{{\bf{r}}}(\mathcal{F}_{n}\frac{j_{i}j_{j}}{\rho})_{{\bf{r}}}\partial_{m}^{r'}\de\brr
\end{equation}

$$\rho({\bf{r}})=\rho_c ({\bf{r}})+ \frac{1}{2}\epsilon_{ijk}\Omega_{k}\pa_{i}j_{(c)j}({\bf{r}})$$
$$j_{i}({\bf{r}})= j_{(c)i}({\bf{r}}) +\rho_{0}\int dx [ \frac{1}{2}\epsilon_{ljk}\Omega_{k}\dot{q_{i}}\dot{q_{j}}\pa_{l}^{{\bf{r}}}\delta({\bf{q(x)-r}}) +\frac{e}{2}\epsilon_{ilk}q_{l}B_{k}\delta({\bf{q(x)-r}}) ]$$
\vskip .3cm
$\int d^3x ~j_{i}=\int d^3x~\rho v_{i}$ as translation generator in  absence of  external magnetic field ${\bf{B}}$:\\
 (i) Let us compute the Poisson bracket  $\lbrace \alpha(\rho) , \rho v_{l} \rbrace$  using eq.s (3,4,5) from the main text,  with ${\bf{B}}=0$.

Straightforward calculation yields
		\begin{equation}
	\label{roji3}
    \lbrace \alpha({\bf{r}}) ,\int d^3 r' \rho v_{l} ({\bf{r'}})  \rbrace =\frac{\pa\alpha}{\pa \rho} ({\bf{r}}) \left[\int d^3 r' \epsilon_{mjk}\Omega_{k}\pa_{m}^{{\bf{r}}}(\rho v_{l}\pa_{j}^{{\bf{r}}}\de  \brr) + \rho ({\bf{r'}})\pa_{l}^{{\bf{r}}}\delta\brr\right]
\end{equation}
We drop the first term as it is a total derivative term to obtain
	\begin{equation}
	\label{rojifinal}
    \lbrace \alpha({\bf{r}}) ,\int d^3 r' \rho v_{l}({\bf{r'}}) \rbrace =\frac{\pa\alpha}{\pa \rho}({\bf{r}}) \left[\int d^3 r'  \rho ({\bf{r'}})\pa_{l}^{{\bf{r}}}\delta\brr\right] = \frac{\pa\alpha}{\pa \rho} \pa_{l}\rho({\bf{r}}) = \pa_{l}\alpha({\bf{r}}) .
\end{equation}
	
	Next, we consider an arbitrary function of  velocity $A_{l}(v_{i})$.
 After some non trivial algebra we find, using
 \begin{equation}
 \label{vjj4}
  \lbrace v_{i}({\bf{r}}) , \int d^3 r' \rho v_{j}({\bf{r'}}) \rbrace = \int d^3 r' [\delta \brr \pa_{j}v_{i} - v_{i}({\bf{r}})\pa_{j}^{{\bf{r'}}}\delta\brr] =  \pa_{j}v_{i} ({\bf{r}})
\end{equation}
the final result
	\begin{equation}
	    \label{vjifinal}
	     \lbrace A_{l}({\bf{r}}) ,\int d^3 r' \rho v_{j} ({\bf{r'}})\rbrace = \frac{d A_{l}}{d v_{i}}  \pa_{j}v_{i}({\bf{r}})  = \pa_{j} A_{l}({\bf{r}}).
	\end{equation}